\documentclass[%
 reprint,
 amsmath,amssymb,
 aps,
pra,
]{revtex4-1}

\usepackage{graphicx}% Include figure files
\usepackage{dcolumn}% Align table columns on decimal point
\usepackage{bm}% bold math
\begin{document}
\preprint{APS/123-QED}
\title{Autler-Townes doublet in single-photon Rydberg spectra \\
of Cesium atomic vapor with a 319 nm UV laser}% Force line breaks with \\
\author{Jiandong Bai${}^{1}$}
\author{Jieying Wang${}^{1}$}
\author{Shuo Liu${}^{1}$}
\author{Jun He${}^{1,2}$}
\author{Junmin Wang${}^{1,2,}$}
\email{wwjjmm@sxu.edu.cn}
\affiliation{{${}^{1}$State Key Laboratory of Quantum Optics and Quantum Optics Devices, Institute of Opto-Electronics, Shanxi University, Taiyuan 030006, China}\\{${}^{2}$Collaborative Innovation Center of Extreme Optics, Shanxi University, Taiyuan 030006, China}}
\begin{abstract}
We demonstrate the single-photon excitation spectra of cesium Rydberg atoms by means of a Doppler-free purely all-optical detection with a room-temperature vapor cell and a 319 nm ultra-violet (UV) laser. We excite atoms directly from ${{6S}_{1/2}}$ ground state to ${{71P}_{3/2}}$ Rydberg state with a narrow-linewidth 319 nm UV laser. The detection of Rydberg states is performed by monitoring the absorption of an 852 nm probe beam in a V-type three-level system. With a strong coupling light, we observe the Autler-Townes doublet and investigate experimentally the dependence of the separation and linewidth on the coupling intensity, which is consistent with the prediction based on the dressed state theory. We further investigate the Rydberg spectra with an external magnetic field. The existence of non-degenerate Zeeman sub-levels results in the broadening and shift of the spectra. It has potential application in sensing magnetic field.
\end{abstract}

\maketitle

%\tableofcontents

%\section{Introduction}
\section{Introduction}
\textit{}Rydberg atoms, highly-excited atoms with principal quantum numbers \emph{n} $\gg$ 1, have shown great promise in terms of their unique structure. These physical properties scale with principal quantum number \emph{n}, such as long radiation lifetime ($\sim$${{n}^{3}}$), strong dipole-dipole interaction ($\sim$${{n}^{4}}$) and large polarizability ($\sim$${{n}^{7}}$) [1]. It makes Rydberg atoms promising for applications such as quantum information processing [2], quantum metrology [3, 4] and simulations of quantum many-body physics [5]. The orbital radius ($\sim$${{n}^{2}}$) of Rydberg atoms is very large, and the binding energy ($\sim$${{n}^{-2}}$) decreases with principal quantum number, making it susceptible to interference from an external field (electric field, magnetic field and microwave field). A new quantum interference phenomenon will be generated by combining the sensitivity of the external field and the quantum coherence effect. Autler-Townes (AT) splitting [6] and electromagnetically induced transparency (EIT) [1, 7] are typical quantum interference effects, and have been widely investigated experimentally and theoretically [8-10]. These effects can be applied to study various physical phenomena, such as four-wave mixing [11], slow light [12], dephasing rates of Rydberg states [13], enhancement of spin-orbit interaction [14], and the measurement of the interval of atomic hyperfine levels. Moreover, the AT splitting can be also used to measure the Rabi frequency [15]. To explore the implication of AT splitting, a weak probe light is scanned through a thermal atomic vapor cell, which is pumped by an intense coupling light, to observe a probe transmission spectrum. However, it is inevitable that the spectra always contain a Doppler background due to the motion of thermal atoms in the vapor cell, and the Doppler broadening greatly limits the resolution of the spectra. The cold atomic ensemble is an ideal medium for the Doppler-free spectra, but the complexity of the laser system results in its failure to be widely used. So the sub-Doppler and Doppler-free technologies in thermal atomic system are still the object many physicists pursue in this field. Rapol \emph{et al.} [16] demonstrated the Doppler-free spectroscopy in driven three-level systems. Wang \emph{et al.} [17] proposed a novel sub-Doppler technology in a Cs vapor cell. Later, Thoumany \emph{et al.} [18] reported Doppler-free single-photon Rydberg excitation spectroscopy with a 297 nm ultra-violet (UV) laser in a room-temperature Rb atomic vapor cell. At present, most of the applications based on the quantum interference effect of Rydberg atoms are mainly focused on the thermal atomic vapor.

In addition, combing the sensitivity of Rydberg atoms to the external field and a three-level system, high-precision Rydberg spectra related to the Zeeman effect play an important role in quantum information processing [19]. In 2016, Bao \emph{et al.} [20] demonstrated the importance of the laser-field polarizations in Rydberg EIT, and modeled qualitatively the Zeeman spectra using standard density matrix equations. Level populations and coherences in ladder-type Rydberg EIT systems are mainly affected by the Zeeman shifts. Recently, Cheng \emph{et al.} [21] observed the phenomenon of $\Lambda$-type EIT in a magnetic field at the room-temperature ${}^{87}$Rb vapor with a buffer gas. Considering the extraordinary characteristics of Rydberg atoms, it is meaningful in a precision magnetometer.

In this paper, using the Doppler-free purely all-optical detection method, we obtain the single-photon Rydberg excitation spectra for ${{6S}_{1/2}}(F = 4)\rightarrow{{71P}_{3/2}}$ transition in a room-temperature Cs atomic vapor cell. In the V-type three-level system, the probe and coupling lasers couple the same ground level to different excited levels: first excited state and Rydberg state. By scanning the strong coupling laser with the locked probe laser, we observe Doppler-free spectra and predict features for the dressed states created by the coupling laser. The separation and linewidth of the AT doublet depend on the coupling intensity as predicted from a simple three-level system. Furthermore, the Rydberg spectra are observed when an external magnetic field is applied. The splitting of Rydberg spectra induced by the non-degenerate Zeeman sub-levels is analyzed.

% FIG. 1
\begin{figure}[tbp]
\vspace{0.00in}
\centerline{
\includegraphics[width = 73mm]{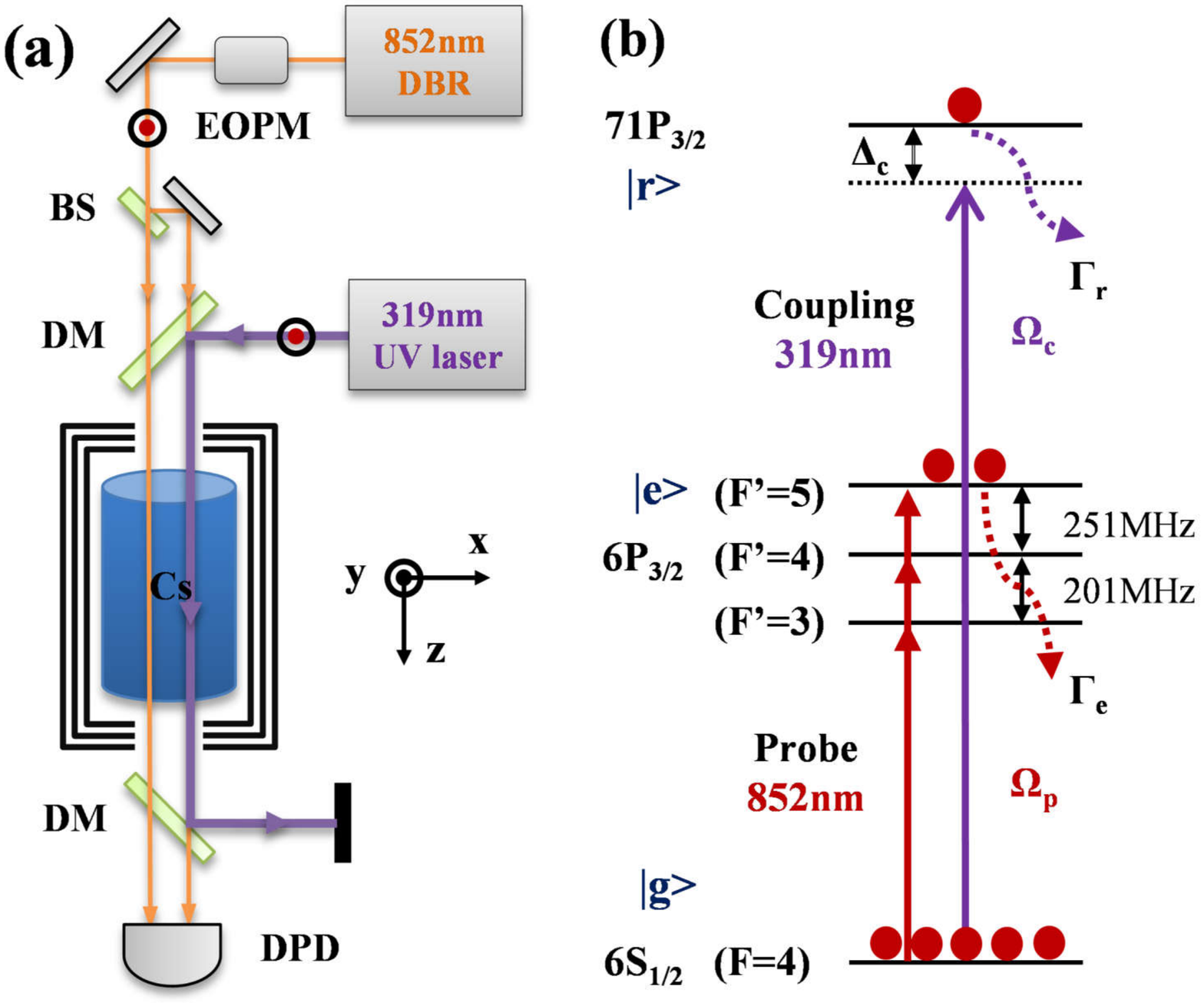}
} \vspace{-0.17in} %\setlength{\columnwidth}{3.2in}
\caption{(a) Schematic of the experimental arrangement. The 319 nm coupling beam co-propagates with the 852 nm probe beam in a 10-cm-long Cs vapor cell placed in a ¦Ì-metal tank. DBR, distributed-Bragg-reflector diode laser; EOPM, electro-optic phase modulator; BS, beam splitter; DM, dichroic mirror; DPD, differential photodiode. (b) Schematic diagram of a V-type three-level system for Cs atomic single-photon Rydberg excitation. The coupling and probe beams have Rabi frequency ${{\Omega}_{c}}$ and ${{\Omega}_{p}}$, respectively. The ${{\Gamma}_{i}}$ (\emph{i} = r or e) denote the decay rates from the respective levels. The 852 nm probe beam is resonant on the transition from ${{6S}_{1/2}}$(F = 4) to ${{6P}_{3/2}}$($F'$ = 3, 4, or 5), and the 319 nm coupling beam is scanned over ${{6S}_{1/2}}(F = 4)\rightarrow{{71P}_{3/2}}$ Rydberg transition.}
\label{Fig 1}
\vspace{-0.2in}
\end{figure}

\section{Experimental arrangement}

The lifetime of Rydberg state increases with principal quantum number. Considering vacuum induced radiative decay and blackbody radiation effects [22], the lifetime of Cs ${{71P}_{3/2}}$ state is 185 $\mu s$, resulting in a natural linewidth of $\sim$ $2\pi\times 860$ Hz. The oscillator strength of ${{6S}_{1/2}}\rightarrow{{71P}_{3/2}}$ transition is six orders of magnitude smaller than that of ${{6S}_{1/2}}\rightarrow{{6P}_{3/2}}$ transition. Compared to ${{6P}_{3/2}}$ excited state, the lower transition probability and the smaller light-scattering cross sections lead to very weak absorption signals for ${{71P}_{3/2}}$ Rydberg state. To enhance the signal, the Rydberg transition is detected by monitoring the reduced absorption of the 852 nm probe beam. A schematic diagram of the experimental setup is shown in Fig. 1a. The narrow-linewidth UV laser is produced by the cavity-enhanced second harmonic generation following sum-frequency generation of two infrared lasers at 1560 nm and 1077 nm, leading to more than 2 W output power at 319 nm [23, 24]. The 852 nm probe beam is provided by a distributed-Bragg-reflector (DBR) diode laser which is frequency stabilized to Cs ${{6S}_{1/2}}(F = 4)\rightarrow{{6P}_{3/2}}$ ($F'$ = 5, or 4, or 3) hyperfine transition by means of polarization spectroscopy [25]. A waveguide-type electro-optic phase modulator (EOPM) (Photline, NIR-MX800) at 852 nm is used to calibrate the frequency interval. The probe beam has a ${{1/e}^{2}}$ radius of $\sim$350 $\mu m$ while the coupling beam is slightly larger with a ${{1/e}^{2}}$ radius of $\sim$400 $\mu m$. The probe light is split into two beams of equal intensity via a beam splitter. One of them is superposed with the 319 nm coupling beam using a dichroic mirror, and co-propagate through a 10-cm-long fused-quartz Cs cell placed in a $\mu$-metal tank. This tank reduces the magnetic field to less than 0.2 mG. To eliminate the Doppler broadened background, the transmission of the probe beams are detected with a differential photodiode (DPD) (New Focus, Model 2107) after passing through another dichroic mirror. When the UV light is scanned across a specific single-photon Rydberg transition, the atoms on ground state are partially transferred to the long-lifetime Rydberg state, leading to the reduced absorption of the probe beam overlapped with the coupling beam.

Figure 1b shows a V-type three-level system interacting with two light fields composed of the ground state $\mid$g$\rangle$ [${{6S}_{1/2}}(F = 4)$], the excited state $\mid$e$\rangle$ [${{6P}_{3/2}}$($F'$ = 3, 4, 5)], and a Rydberg state $\mid$r$\rangle$ [${{71P}_{3/2}}$]. For the V-type system, the common ground $\mid$g$\rangle$ is coupled to two excited states: $\mid$g$\rangle$$\leftrightarrow\mid$e$\rangle$ transition by the probe light and $\mid$g$\rangle$$\leftrightarrow\mid$r$\rangle$ transition by the UV coupling light. The coupling beam has Rabi frequency of ${{\Omega}_{c}}$ and detuning of ${{\Delta}_{c}}$ from resonance. The probe beam has Rabi frequency ${{\Omega}_{p}}$ and detuning ${{\Delta}_{p}}$. The decay rates from the excited state $\mid$e$\rangle$ and Rydberg state $\mid$r$\rangle$ are ${{\Gamma}_{e}}$ and ${{\Gamma}_{r}}$, respectively. Here ${{\Gamma}_{e}}$ is $2\pi\times 5.2$ MHz.

If the probe light is locked on ${{6S}_{1/2}}(F = 4)\rightarrow{{6P}_{3/2}}(F' = 5)$ cycling transition, the atoms with zero-velocity component along the propagation direction of the probe beam populate in the ${{6P}_{3/2}}(F' = 5)$ state, as shown in Fig. 1b. Considering the Doppler effect in a thermal ensemble, the ${{6P}_{3/2}}$($F'$ = 4 and 3) states are also populated atoms with a certain velocity component in the same direction. For the atoms with a velocity component of ${{\upsilon}_{z}}$, the detuning of probe and coupling fields in the same direction are expressed as:
%Eqn{1}
\begin{eqnarray}
\Delta_p &=& k_p \upsilon_z = -\frac{\omega_0}{c} \upsilon_z\nonumber\\
\Delta_c &=& k_c \upsilon_z = -\frac{\lambda_p}{\lambda_c} \upsilon_z
\end{eqnarray}

Where ${{\omega}_{0}}$ is the reference frequency; c is the speed of light in vacuum; ${{\lambda}_{p}}$, ${{\lambda}_{c}}$ are the wavelengths of the probe and coupling beams, respectively.

\section{Results and discussion}
\subsection{Single-photon Rydberg excitation velocity-selective spectra}

In general, the interaction between the atoms and the coupling light field is studied by the absorption of probe beam. In this way, we can obtain the coherent spectra with Doppler-free background, and also improve the spectral resolution. Since the frequency interval between ${{6P}_{3/2}}(F' = 3)$ and $(F' = 4)$ is 201 MHz which is much larger than the laser linewidth of 1 MHz, and the probe light is locked to the ${{6S}_{1/2}}(F = 4)\rightarrow{{6P}_{3/2}}(F' = 5)$ cycling transition, the absorption of the probe beam attributes mainly to the atoms with velocity of ${{\upsilon}_{z}}$ = 0 in the direction of beam propagation. When the probe light with a certain intensity pass through Cs vapor cell, it will immediately reach a population balance between the ground state and ${{6P}_{3/2}}$ excited state, and a stable transmission signal is obtained simultaneously. Only when some atoms are transferred from the ground state to the ${{71P}_{3/2}}$ Rydberg state, the balance between the ground state and ${{6P}_{3/2}}$ excited state is broken, and the absorption of the probe light is reduced. Therefore, the absorption spectra of single-photon Rydberg excitation is observed by the enhanced transmission of the probe beam while scanning the frequency of the coupling beam.

% FIG. 2
\begin{figure}[tbp]
\setlength{\belowcaptionskip}{-0.5cm}
\vspace{-0.05 in}
\centerline{
\includegraphics[width=100mm]{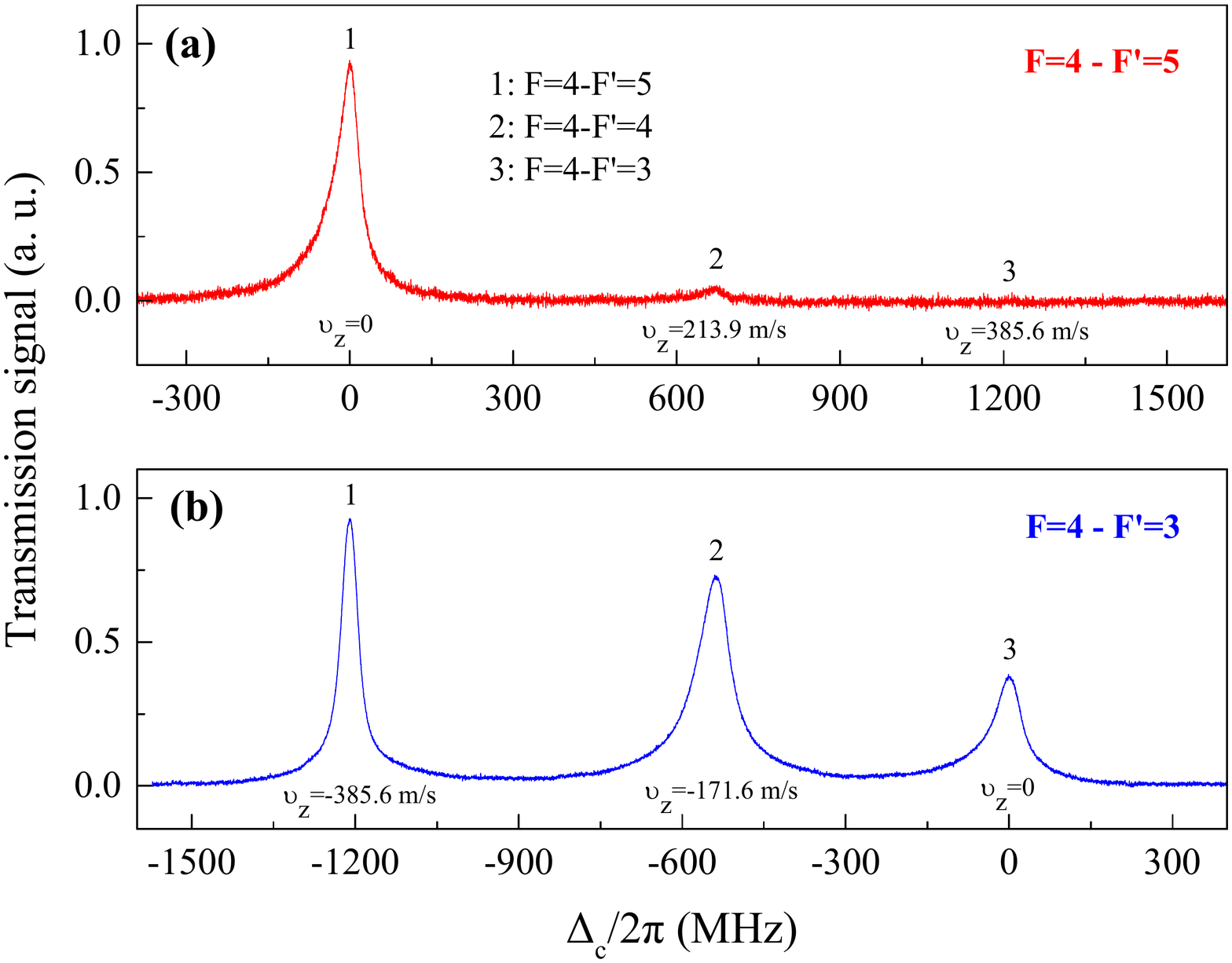}
} \vspace{-0.15in} %\setlength{\columnwidth}{0.15in}
%\centerline{
%}
\caption{The single-photon Rydberg excitation velocity-selective spectra for the ${{6S}_{1/2}}(F = 4)\rightarrow{{71P}_{3/2}}$ transition when the probe light is locked to different transitions: (a) Cs ${{6S}_{1/2}}(F = 4)\rightarrow{{6P}_{3/2}}(F' = 5)$ hyperfine transition, (b) Cs ${{6S}_{1/2}}(F = 4)\rightarrow{{6P}_{3/2}}(F' = 3)$ hyperfine transition.}
\label{Fig 2}
\vspace{-0.01in}
\end{figure}

Figure 2 shows the single-photon excitation spectra for the ${{71P}_{3/2}}$ Rydberg state when the probe light is locked to different hyperfine transitions from Cs ${{6S}_{1/2}}(F = 4)$ to ${{6P}_{3/2}}$. The probe and coupling beams have identical linear polarization. Figure 2a, b correspond to the probe light are resonant with ${{6S}_{1/2}}(F = 4)\rightarrow{{6P}_{3/2}}(F' = 5)$ and ${{6S}_{1/2}}(F = 4)\rightarrow{{6P}_{3/2}}(F' = 3)$ transitions. When the UV light is scanned over the ${{71P}_{3/2}}$ state, three transmission peaks (marked as 1, 2, 3) appear and correspond to $(F = 4)\rightarrow(F' = 5)$, $(F = 4)\rightarrow(F' = 4)$, and $(F = 4)\rightarrow(F' = 3)$ hyperfine transitions. Here, we use an EOPM to produce two sidebands to calibrate the frequency interval of the transmission spectra. Setting ${{6S}_{1/2}}(F = 4)\rightarrow{{6P}_{3/2}}(F' = 5)$ cycling transition as the reference frequency, for atoms with velocity of ${{\upsilon}_{z}}$ = 0, the detuning of the probe light is ${{\Delta}_{p}}$ = 0; for atoms with velocity groups of ${{\upsilon}_{z}}$ = 213.9 and 385.6 m/s, ${{\Delta}_{p}}$ are -251 and -452 MHz given by Eq. (1), which will result in resonance with $(F = 4)\rightarrow(F' = 4)$ and $(F = 4)\rightarrow(F' = 3)$ hyperfine transitions, respectively. Considering the wavelength mismatch of ${{\lambda}_{p}}/{{\lambda}_{c}}$=2.675 between the probe and coupling lasers, when the UV laser matches these atomic groups to the ${{71P}_{3/2}}$ state, the corresponding detuning ${{\Delta}_{c}}$ are 671 and 1210 MHz, which are consistent with the measurement values with a high-precision wavelength meter (HighFinesse, WS-7), as shown in Fig. 2a. For Fig. 2b, the interacting process is also similar and it does not vary with different ${{nP}_{3/2}}$ Rydberg states (\emph{n}=70-100).

Furthermore, we have observed that the relative intensity of the three transmission peaks (marked as 1-3) is different when the probe light is resonant with different atomic transitions. There are two main reasons: the strength of the interaction between atoms and nearly-resonant laser, and the atoms with a certain velocity interacting with lasers. The strength of the interaction between two Zeeman sub-levels is characterized by the dipole matrix elements [26]:
%Eqn{2}
\begin{eqnarray}
\lefteqn{\langle JIFm_F\mid er_q\mid J'I'F'm_{F'}\rangle}\nonumber\\ &=& \langle J\Vert er_q\Vert J'\rangle (-1)^{2F'+m_F+J+I}\nonumber\\ && \times\sqrt{(2F+1)(2F'+1)(2J+1)}\nonumber\\ &&
\times\begin{pmatrix} F' & 1 & F\\ {{m}_{F'}} & q & -{{m}_{F}}\end{pmatrix}
\begin{Bmatrix} J & J' & 1\\ F' & F & I\end{Bmatrix}
\end{eqnarray}

Here, $q={{m}_{F'}}-{{m}_{F}}$ is polarization dependent, labeling the component of \emph{r} in the spherical basis, and the doubled bars indicate that the matrix element is reduced. \emph{I} is the nuclear spin momentum, \emph{J} is the angular momentum, \emph{F} is the total angular momentum, and ${{m}_{F}}$ is the Zeeman sub-levels. The values in the parentheses and curly brackets denote the Wingner 3-j symbol and Wingner 6-j symbol, respectively. Due to the two beams have identical linear polarization, we only calculate the relative transition strengths between two Zeeman sub-levels for $\pi$ transitions $[(F = 4, {{m}_{F}})\rightarrow(F' = 5, 4, 3, {{m}_{F'}}={{m}_{F}})]$, the corresponding ratio of the three transmission peaks should be 44: 21: 7. In fact, due to the laser interacts with the atoms of different velocities, the relative strength of these transmission peaks does not perfectly satisfy this condition.

When the physical system is in thermal equilibrium, most of the atoms are in the ground state, and its velocity distribution is described by the Maxwell-Boltzmann distribution. Taking the transmission peak (marked as 3) as an example, when the detuning of the UV laser is relatively small, the transmission enhancement of the signal becomes more obvious with the increasing of the number of atoms interacting with the laser in the direction of beam propagation. On the contrary, for the peak-1 in the red detuning, the relative strength of the transmission signal will gradually decrease with the increase of the detuning, as shown in Fig. 2.

\subsection{Aulter-Towns splitting}

We further investigate the transmission spectra for different coupling intensities with the probe light of $\sim$43 $\mu$W, as shown in Fig. 3. The 852 nm probe light is locked to the hyperfine transition of ${{6S}_{1/2}}(F = 4)\rightarrow{{6P}_{3/2}}(F' = 3)$. The transmission peak (marked as 3) corresponds to ${{6S}_{1/2}}(F = 4)\rightarrow{{6P}_{3/2}}(F' = 3)$ hyperfine transitions, as depicted in Fig. 2b. With a moderate intensity of the coupling light, we can see that the AT doublet appears in the transmission spectra. The separation of the doublet increases with the enhancement of the coupling light, and its linewidth is also gradually widened.

% FIG. 3
\begin{figure}[tbp]
\setlength{\belowcaptionskip}{-0.5cm}
\vspace{-0.05in}
\centerline{
\includegraphics[width=100mm]{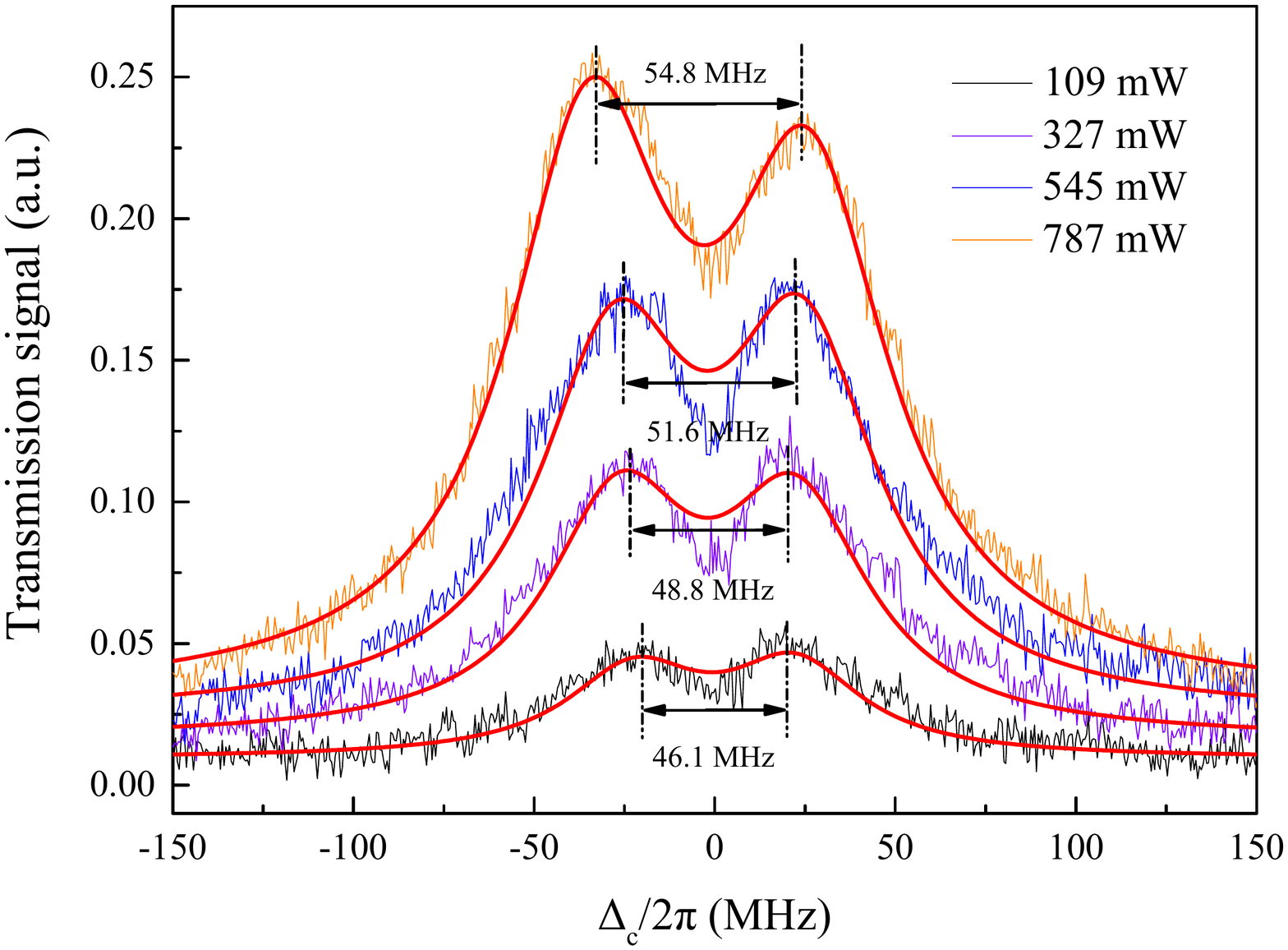}
} \vspace{-0.25in} %\setlength{\columnwidth}{3.2in}
%\centerline{
%}
\caption{The AT splitting spectra at different coupling intensities for the certain intensity of probe light. The probe light is locked to the hyperfine transition of  ${{6S}_{1/2}}(F = 4)\rightarrow{{6P}_{3/2}}(F' = 3)$. The transmission peak corresponds to Cs  ${{6S}_{1/2}}(F = 4)\rightarrow{{6P}_{3/2}}(F' = 3)$ hyperfine transition. The red solid lines is the results of the mutiple-peaks fitting of Lorentz function.}
\label{Fig 3}
\vspace{0.1in}
\end{figure}

For the above physical process, we first consider a simple V-type three-level system, as shown in Fig. 1b. When the strong coupling light interacting with a two-level system ($\mid$g$\rangle$$\leftrightarrow\mid$r$\rangle$ transition), the atomic energy level is split into two which is induced by dressing splitting, resulting in a new resonance. The absorption of the probe beam is proportional to Im (${{\rho}_{eg}}$), where ${{\rho}_{eg}}$ is the induced polarizability on the $\mid$e$\rangle$$\leftrightarrow\mid$g$\rangle$ transition coupled by the probe light. From the density-matrix equations, the steady-state value of ${{\rho}_{eg}}$ is given by [27]:
%Eqn{3}
\begin{eqnarray}
\rho_{eg} &=& \frac{i\Omega_p(\gamma_{rg}\Gamma_{rg}+2{\Omega_c}^2)(i\Delta_p+\gamma_{eg})}
{[(i\Delta_p+\Gamma_{eg})(i\Delta_p+\Gamma_{re})+{\Omega_c}^2](\gamma_{rg}\Gamma_{rg}+4{\Omega_c}^2)}
\end{eqnarray}

Where ${{\gamma}_{ij}}$ and ${{\Gamma}_{ij}}$ denote the spontaneous emission rates and the decay rates from $\mid$\emph{i}$\rangle$ to $\mid$\emph{j}$\rangle$, respectively. In our case, the $\mid$r$\rangle$$\leftrightarrow\mid$e$\rangle$ transition is dipole-forbidden, and ${{\Gamma}_{re}}$ represents higher-order transition. The theoretical analysis process are similar to that of the $\Lambda$-system [10], showing that the absorption spectra of probe beams are separated into two peaks located at the position of two dressed states, named the AT doublet. Their locations are given by [28]:
%Eqn{4}
\begin{eqnarray}
\Delta_\pm &=& \frac{\Delta_c}{2}\pm\frac{1}{2}\sqrt{{\Delta_c}^2+4{\Omega_c}^2}
\end{eqnarray}

\leftline{with the corresponding linewidths:}
%Eqn{5}
\begin{eqnarray}
\Gamma_\pm &=& \frac{\Gamma_{rg}+D}{2}(1\mp\frac{\Delta_c}{\sqrt{{\Delta_c}^2+4{\Omega_c}^2}})+W
\end{eqnarray}

Here, $D$ is the Doppler width and $W$ is the other broadenings analyzed in the following section. It is clear from the above expression that, if ${{\Delta}_{c}}\approx$ 0, the location of the two peaks have ${{\Omega}_{c}}$ symmetrically shifted away from the atomic hyperfine transition and they have identical linewidth of (${{\Gamma}_{rg}}+D$)/2. In this situation, the separation between the AT doublet uniquely depends on the coupling Rabi frequency. In our experiment, the probe light is locked to Cs ${{6S}_{1/2}}(F = 4)\rightarrow{{6P}_{3/2}}(F' = 3)$ transition, so the transmission signal obtained with DPD is mainly caused by zero velocity atoms which have a certain probability to absorb the coupling and probe beams.

% FIG. 4
\begin{figure}[tbp]
\vspace{0.05in}
\centerline{
\includegraphics[width=90mm]{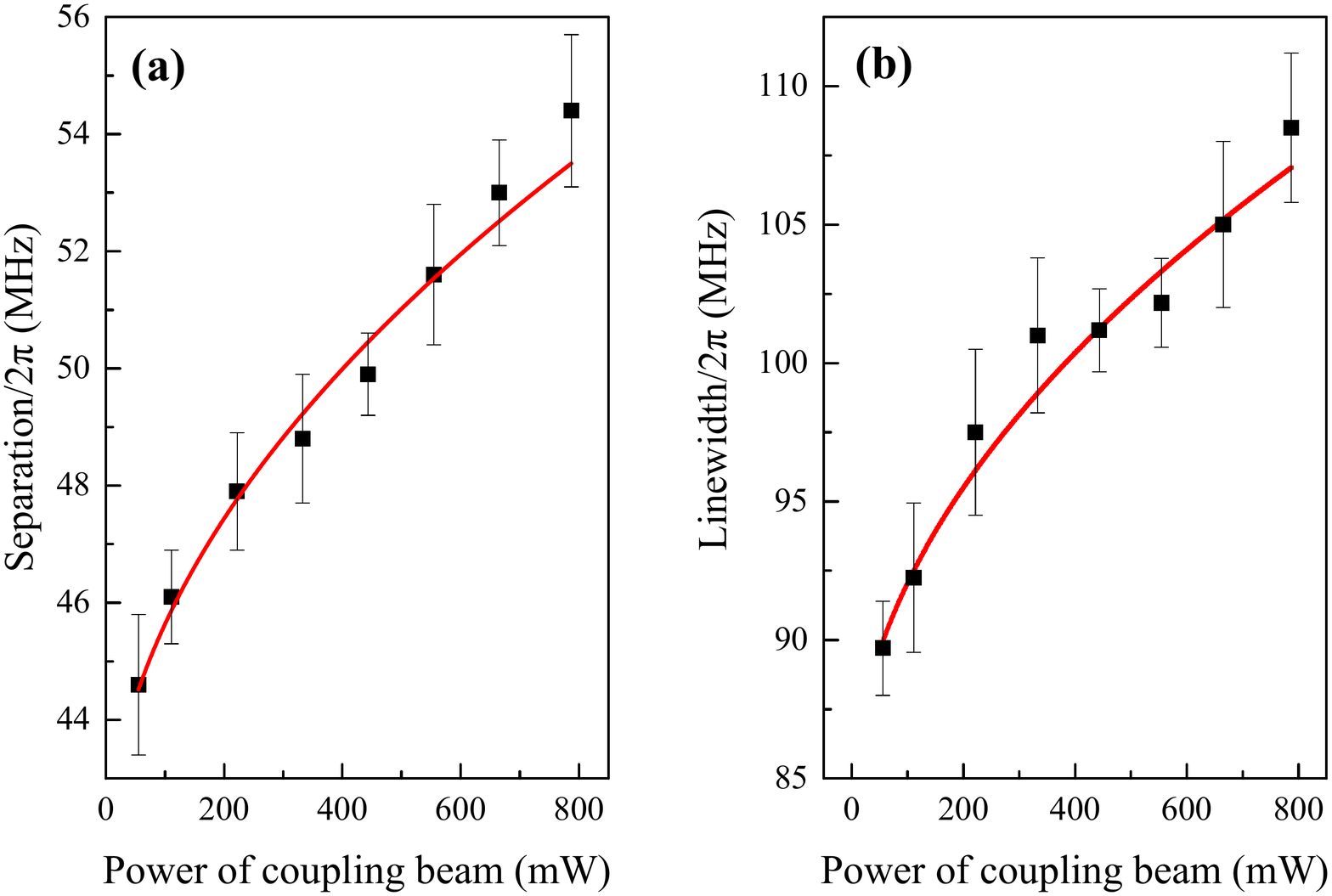}
} \vspace{-0.2in} %\setlength{\columnwidth}{3.2in}
%\centerline{%}
\caption{ The separation and linewidth of the peak-3 in the AT doublet of single-photon Rydberg spectra as a function of the coupling beam's power, while the probe light is resonant with Cs  ${{6S}_{1/2}}(F = 4)\rightarrow{{6P}_{3/2}}(F' = 3)$ transition. The solid lines are the expected variation from the theoretical analysis, as explained in the text. The error bar is obtained by fitting the AT doublet, including the fluctuation of the laser beams and the asymmetry of the spectrum due to the nonlinear sweep of the triangular wave for the UV laser.}
\label{Fig 4}
\vspace{-0.15in}
\end{figure}

Figure 4 shows that the separation and linewidth of the AT doublet as a function of the coupling beam's power. The relation between the effective Rabi frequency and the power depends on many factors such as the intensity profile of the coupling beam, absorption through the cell, scattering loss at the surface of the cell. Moreover, the measured lifetime of the Rydberg state is much shorter than the calculated value due to atom loss induced by an external decoherence mechanism [1, 22], such as laser irradiation of surfaces. It is difficult to determine experimentally these factors precisely, so we have left it as an overall fit parameter to obtain the solid lines in Fig. 4. It is important to note that this parameter does not change the trend of the data and its shape, but only makes the fitted curve to move up or down. The best fitting results show that the coupling beam has an ideal Gauss radius of $\sim$430 $\mu$m, which is slightly larger than the measured value of $\sim$400 $\mu$m. Considering the experimental condition of ${{\Delta}_{c}}\approx$ 0 and the existing broadening mechanisms described below in the text, the solid line in Fig. 4a is only the variation predicted by Eq. (4). The trend of experimental data is in good agreement with the theoretical prediction. In Fig. 4b, with the fixed intensity of the probe laser, the linewidth of the AT doublet increases with the coupling beam's power, which is mainly caused by the power broadening. The power-broadened linewidth varies as $\omega={{\Gamma}_{Nat}}\sqrt{1+s}$, where $s=I/{{I}_{s}}$ is the saturation parameter. Taking into account various complex broadening mechanisms, the expression can be written as:
%Eqn{6}
\begin{eqnarray}
\omega &=& \Gamma_{Nat}\sqrt{1+s}+W
\end{eqnarray}

From the fitting results shown in Fig. 4b, we estimate that the measured narrowest linewidth is about 83.8 MHz (not zero) without the coupling beam. We first attribute the remaining linewidth to power broadening by the probe beam. Considering the wavelength mismatch of ${{\lambda}_{p}}/{{\lambda}_{c}}$=2.675 between the probe and coupling beams, we calculate that the power broadened linewidth of probe beam for one of the doublets is 32.2 MHz, which is smaller than the measured value. Therefore, there exist other broadening mechanisms in the V-type Rydberg system, not only the power broadening. It also contributes from the following broadening mechanisms: natural linewidth of $\sim$14 MHz ($\sim2.675\times$5.2 MHz) [29], transit-time broadening of $\sim$450 kHz ($\sim2.675\times$168 kHz), collisional broadening [30], misalignment of the probe and coupling beams in Cs atomic vapor cell [16], the long-range interactions between Rydberg atoms, and the broadening caused by stray electronic fields. The fact that an applied dc electric field is not effective on the EIT spectra regardless of the probe and coupling laser polarizations due to Rydberg atoms are being screened in the vapor cell [31]. The shielding effect is attributed to ions and electrons produced by photodesorption induced by the coupling laser at the cell inner surfaces. In the presence of an additional radio-frequency field, the shielding effect persists and is reduced by an external high-frequency field (higher than dozens of MHz) [32]. Therefore, the spectral lines broadening from the Stark shifts with the ambient static and low-frequency alternating electric field is negligible. In order to reduce the interference of high-frequency electric field, we will use the shielding material with good conductivity and being grounded in the future. Considering the above broadening mechanisms, the observed linewidth is well fitted. However, it is important to note that the intensity of the probe light is relatively strong in order to observe the Rydberg spectra with high signal-to-noise ratio. Subsequently, we need to improve the signal-to-noise ratio of the signal and suppress spectral broadening of the weak excitation signal which is critical for precision spectroscopy.

\subsection{Rydberg spectra in an external magnetic field}

Hyperfine structure is a result of the electron-nucleus interactions. Each of the hyperfine energy levels contains 2\emph{F}+1 magnetic sub-levels that determines the angular distribution of the electron wave function. Without an external magnetic field, these Zeeman sub-levels are degenerate. However, when the magnetic field is applied, their degeneracy is broken. The total Hamiltonian of the system can be expressed as [26]:
%Eqn{7}
\begin{eqnarray}
H &=& H_{hfs}+H_B
\end{eqnarray}

Using the first-order perturbation theory, the Hamiltonian of hyperfine structure and that of describing the interaction of atoms with magnetic fields are expressed as, respectively:
%Eqn{8}
\begin{eqnarray}
H_{hfs} &=& A_{hfs}\bm{IJ}\nonumber\\&&+B_{hfs}\frac{3(\bm{IJ})^2+\frac{3}{2}(\bm{IJ})-I(I+1)J(J+1)}
{2I(2I-1)J(2J-1)}\nonumber\\
H_B &=& \frac{\mu_B}{h}(g_S\bm{S}+g_L\bm{L}+g_I\bm{I})\bm{B}
\end{eqnarray}

Here ${{A}_{hfs}}$ is the magnetic dipole constant, and ${{B}_{hfs}}$ is the electric quadrupole constant. $\bm{I}$, $\bm{J}$ are the total nuclear and electronic angular momentum, respectively. For ${{}^{133}}$Cs, $\bm{I}$=7/2, ${{\mu}_{B}}$/h $\approx$ 1.4 MHz/Gauss is Bohr magneton, ${{g}_{S}}$, ${{g}_{L}}$, ${{g}_{I}}$ are the electron spin, electron orbital and nuclear \emph{g}-factor, respectively.

% FIG. 5
\begin{figure}[tbp]
\vspace{-0.1in}
\setlength{\belowcaptionskip}{-0.5cm}
\vspace{-0.35in}
\centerline{
\includegraphics[width=50mm]{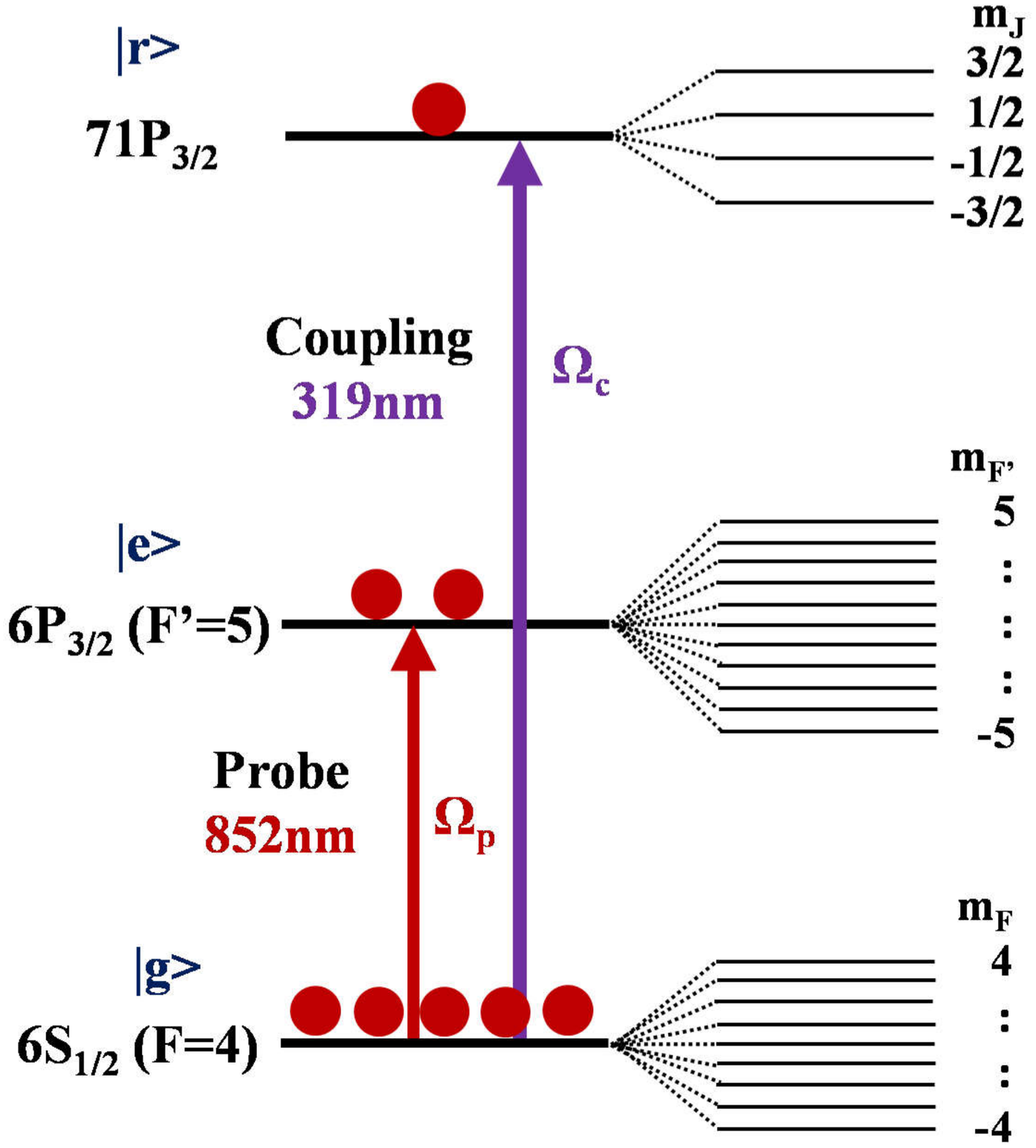}
} \vspace{-0.5in} %\setlength{\columnwidth}{3.2in}
%\centerline{
%}
\caption{The V-type scheme of the single-photon Rydberg excitation without (left) and with (right) magnetic field. The 852 nm probe laser is resonant with the field-free ${{6S}_{1/2}}(F = 4)\rightarrow{{6P}_{3/2}}(F' = 5)$ transition, and the 319 nm coupling laser scans through the ${{6S}_{1/2}}(F = 4)\rightarrow{{71P}_{3/2}}$ Rydberg transition. The sketch on the right shows the Zeeman sub-levels of $\mid$g$\rangle$, $\mid$e$\rangle$ and $\mid$r$\rangle$ states when a magnetic field less than 10 Gauss is applied (separations not to scale).}
\label{Fig 5}
\end{figure}

According to the interaction strength between the atoms and magnetic field, for our V-type three-level system, the splitting of the ground state $\mid$g$\rangle$ [${{6S}_{1/2}}$] is linear Zeeman effect in the weak field, but that of the Rydberg state $\mid$r$\rangle$ [${{71P}_{3/2}}$] is in the strong Paschen-Back region. For the Rydberg state, the decoupled basis ($\mid$$J{{m}_{J}};I{{m}_{I}}$$\rangle$) is used as the eigenbasis while the coupled basis ($\mid$$JIF{{m}_{F}}$$\rangle$) is the eigenbasis for $\mid$g$\rangle$. The dipole matrix elements in the coupled basis have been given in Eq. 2. The decoupled and coupled basis are related by [33]:
%Eqn{9}
\begin{eqnarray}
\mid Jm_J;Im_I\rangle &=& (-1)^{J-I+m_F\sqrt{2F+1}}\nonumber\\ && \times\sum\limits_{Fm_F}
\begin{pmatrix} J & I & F\\ {{m}_{J}} & {{m}_{I}} & -{{m}_{F}}\end{pmatrix}
\mid JIFm_F\rangle
\end{eqnarray}

Using the first-order perturbation theory, the Zeeman splitting of the Rydberg state describing the interaction of atoms with magnetic fields is given by [20, 26]:
%Eqn{10}
\begin{eqnarray}
E_{\mid Jm_J;Im_I\rangle}\nonumber &\approx& A_{hfs}IJ\nonumber\\&&
+B_{hfs}[9(m_I m_J)^2-3J(J+1){m_I}^2\nonumber\\&&-3I(I+1){m_J}^2+I(I+1)J(J+1)]\nonumber\\&&
\div[4J(2J-1)I(2I-1)]\nonumber\\&& +\mu_B(g_J m_J+g_I m_I)B
\end{eqnarray}

From the above Eqs. (8, 10), the shifts of three levels ($\mid$g$\rangle$, $\mid$e$\rangle$, and $\mid$r$\rangle$) are different because of different magnetic interactions, just as described in reference [34]. The absolute position of the spectrum varies with the detuning which is related to Zeeman splitting of different states in an external magnetic field. The Zeeman splitting of the ground state varies linearly with the magnetic field. For Rydberg states in the strong Paschen-Back region, the splitting also linearly depends on the magnetic fields in $\mid$${{m}_{I}}, {{m}_{J}}$$\rangle$  basis; for the atoms in ${{6P}_{3/2}}$ excited state, the magnetic interaction is linear and quadratic Zeeman effect. When the magnetic field is small (B $\leq$ 10 Gauss), it is approximatly in the linear range of Zeeman effect. In the magnetic field, for the ground state (${{\Delta}_{g}}$), first excited state (${{\Delta}_{e}}$) and Rydberg state (${{\Delta}_{r}}$), the interval of adjacent Zeeman sub-levels can be expressed as:
%Eqn{11}
\begin{eqnarray}
\Delta_g &=& \mu_B B g_F m_F = \frac{1}{4}\mu_B B m_F\nonumber\\
\Delta_e &=& \mu_B B g_{F'} m_{F'} = \frac{2}{5}\mu_B B m_{F'}\nonumber\\
\Delta_r &=& \mu_B B g_J m_J = \frac{4}{3}\mu_B B m_J
\end{eqnarray}

% FIG. 6
\begin{figure}[tbp]
\setlength{\belowcaptionskip}{-0.5cm}
\vspace{-0.45in}
\centerline{
\includegraphics[width=90mm]{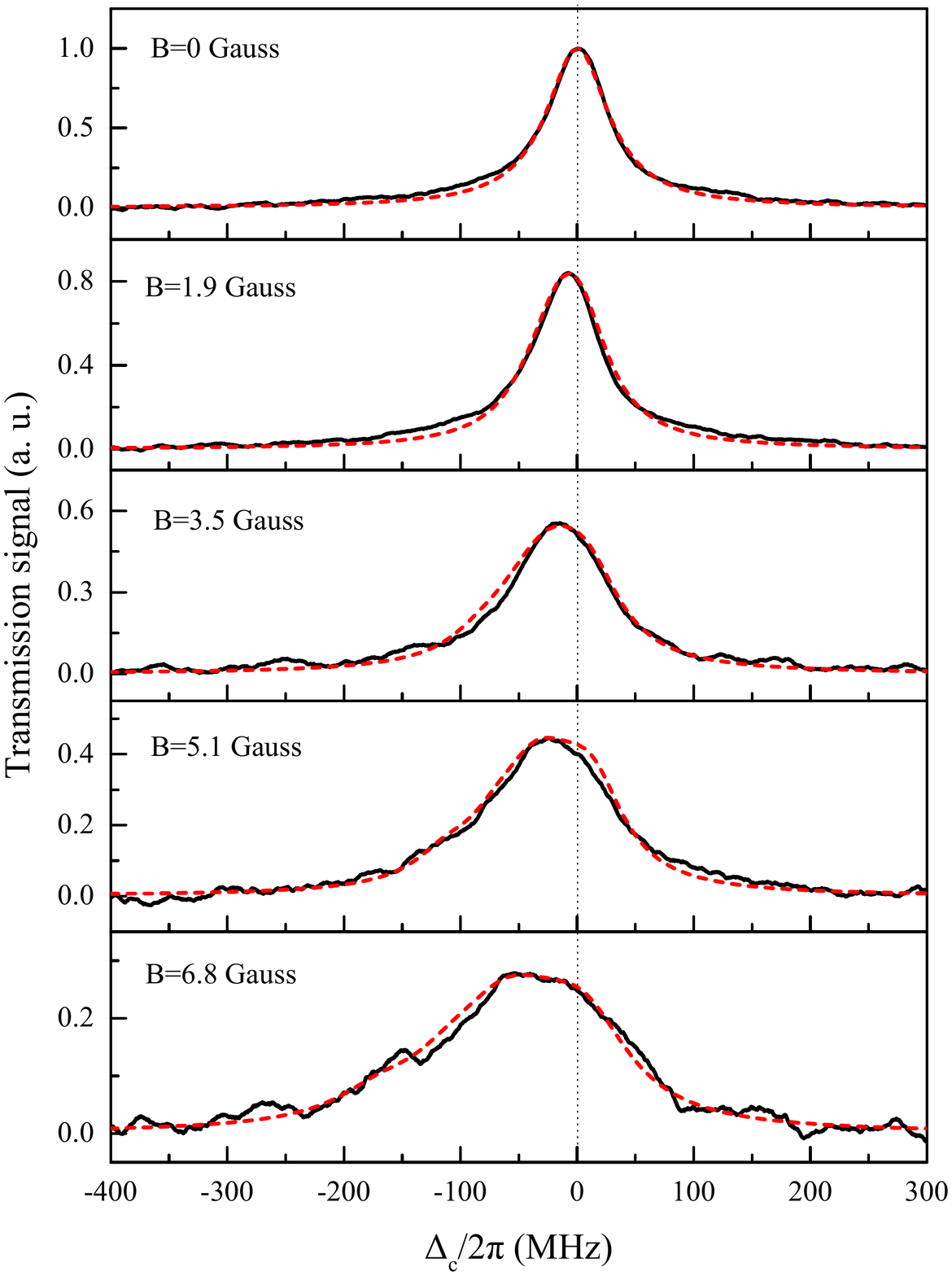}
} \vspace{-0.5in} %\setlength{\columnwidth}{3.2in}
%\centerline{
%}
\caption{The transmission signals of the probe beam in different strengths of magnetic field B from 0 to 6.8 Gauss. The solid lines (black) are the experimental data, the dotted lines (red) are the results of the theoretical fittings. With the increasing of magnetic field, the single-photon Rydberg excitation spectra expand gradually, and the location of the peak is frequency shifted. }
\label{Fig 6}
\end{figure}

% FIG. 7
\begin{figure}[tbp]
\setlength{\belowcaptionskip}{-0.5cm}
\centerline{
\includegraphics[width=90mm]{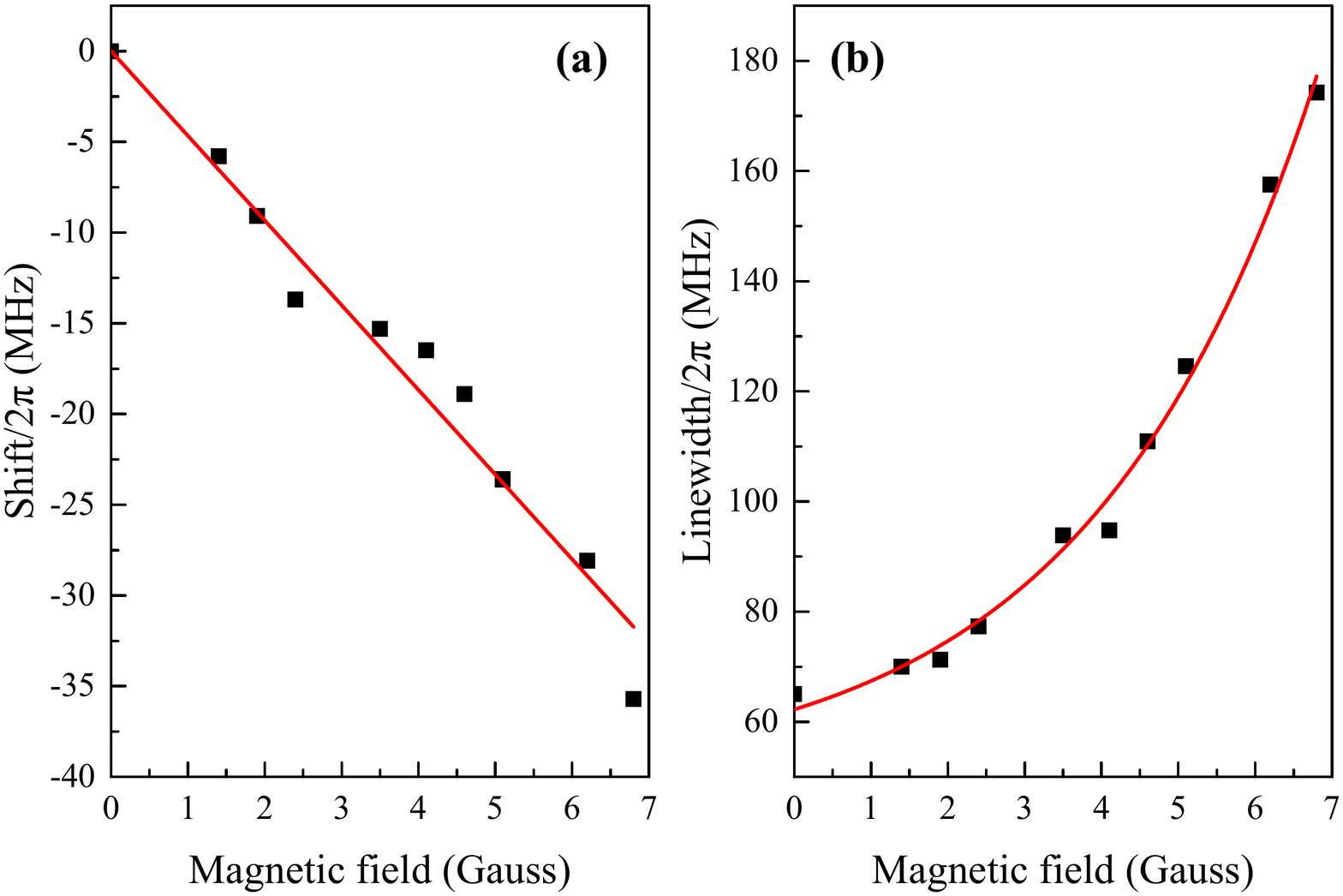}
} \vspace{-0.2in} %\setlength{\columnwidth}{3.2in}
%\centerline{
%}
\caption{The frequency shift and linewidth of the transmission signal as a function of magnetic field. The solid lines are the expected variation from the theoretical prediction.}
\label{Fig 7}
\end{figure}

The detunings of probe and coupling beams in the magnetic field relative to the transition of $\mid$g$\rangle$$\leftrightarrow\mid$e$\rangle$ and $\mid$g$\rangle$$\leftrightarrow\mid$r$\rangle$ are ${{\Delta}_{p}}={{\Delta}_{e}}-{{\Delta}_{g}}$ and ${{\Delta}_{c}}={{\Delta}_{r}}-{{\Delta}_{g}}$, respectively. Considering the wavelength mismatch induced by Doppler effect, when the probe beam is resonant to the field-free ${{6S}_{1/2}}(F = 4)\rightarrow{{6P}_{3/2}}(F' = 5)$ transition, the resonance condition satisfy the relation:
%Eqn{12}
\begin{eqnarray}
\Delta_c &=& (\Delta_r -\Delta_g)+\frac{\lambda_p}{\lambda_c}(\Delta_e -\Delta_g)
\end{eqnarray}

In Eq. (10), the fine energy level of ${{71P}_{3/2}}$ Rydberg state is split into four Zeeman sub-levels of ${{m}_{J}}=\pm1/2, \pm3/2$ in an external magnetic field. The Zeeman splitting of the ${{71P}_{3/2}}$ state is shown in Fig. 5. As shown in Fig. 1a, both the probe and coupling beams are linearly polarized along the y direction, and co-propagate passing through Cs cell, which is contained in a cylindrical solenoid. The coupling and probe beams travel in the direction of the magnetic field (parallel to the z direction). Figure 6 shows that the transmission signals of the probe beam in different strengths of magnetic field B from 0 to 6.8 Gauss. With the increasing of magnetic field, the single-photon Rydberg excitation spectra expand gradually. The transitions between the degenerate Zeeman sub-levels are resonant, and the optical pump effect is perfect. But when an external magnetic field is applied, the degeneracy is broken. The Zeeman splitting is different for $\mid$g$\rangle$, $\mid$e$\rangle$  and $\mid$r$\rangle$ in the same magnetic field, so the transitions between Zeeman sub-levels are no longer resonant. According to the Eq. (11), the detuning increase with the magnetic field, and the corresponding optical pump effect [35] is also gradually weaken. Therefore, the transmission signal decreases with the magnetic field, and the signal is frequency shifted along the direction of red detuning of the coupling beam, as shown in Fig. 7. The solid lines are the expected variation from the theoretical prediction. In previous work [36], due to the influence of power broadening and other broadening mechanisms, the narrowest spectral linewidth is about 24 MHz in the absence of the magnetic field. Hence, in an external magnetic field, we don't observe clearly four Zeeman sub-peaks in the spectra due to the broadened linewidth of the signal limits the spectral resolution. Even so, it has potential application in sensing magnetic field.

\section{Conclusion}

We have demonstrated the Doppler-free velocity-selective spectra of Cs Rydberg atoms with a single-photon excitation scheme in a room-temperature vapor cell. The weak atomic transition from the ground state [${{6S}_{1/2}}$] to Rydberg state [${{71P}_{3/2}}$] is detected by the reduced absorption of probe beam resonant on the ${{D}_{2}}$ line of Cs in a V-type three-level system. With moderate intensity of the coupling light, the Autler-Townes doublet is clearly observed. The dependence of the separation and linewidth on the coupling intensity is experimentally investigated with dressed state model. The results show good agreement with the theoretical prediction. The implement of this experiment opens the possibility to construct the Rydberg-dressed magneto-optical trap in future work, which combines laser cooling with tunable long-range interactions between Rydberg atoms. It could be used for realizing supersolids and spin squeezing for enhanced metrology. Furthermore, we have observed and analyzed the spectra of non-degenerate Zeeman sub-levels when an external magnetic field is applied. Investigation of the single-photon Rydberg spectroscopy in room-temperature atomic vapor has promising application in precision spectroscopy, especially for studying the atomic quantum states and the interaction between Rydberg atoms.

\vspace{-0.15in}
\section{Acknowledgments}

This project is supported by the National Key Research and Development Program of China (2017YFA0304502), the National Natural Science Foundation of China (61475091 and 11774210) and the Shanxi Provincial 1331 Project for Key Subjects Construction.


\begin{thebibliography}{99}

\bibitem{1} M. O. Scully, M. S. Zubairy, Quantum Optics (Cambridge University Press, Cambridge, 1997).
\bibitem{2} M. Saffman, T. G. Walker, K. M{\o}lmer, Quantum information with Rydberg atoms. Rev. Mod. Phys. 82, 2313 (2010).
\bibitem{3} J. A. Sedlacek, A. Schwettmann, H. Kubler, J. P. Shaffer, Atom-based vector microwave electrometry using rubidium Rydberg atoms in a vapor cell. Phys. Rev. Lett. 111, 063001 (2013).
\bibitem{4} S. Kumar, H. Fan, H. K{\"u}bler, A. J. Jahangiri, J. P. Shaffer, Rydberg-atom based radio-frequency electrometry using frequency modulation spectroscopy in room temperature vapor cells. Opt. Express 25, 8625-8637 (2017).
\bibitem{5} I. Lesanovsky, Many-body spin interactions and the ground state of a dense Rydberg lattice gas. Phys. Rev. Lett. 106, 025301 (2011).
\bibitem{6} S. H. Autler, C. H. Townes, Stark effect in rapidly varying fields. Phys. Rev. 100, 703 (1955).
\bibitem{7} M. Fleischhauer, A. Imamoglu, J. P. Marangos, Electromagnetically induced transparency: optics in coherent media. Rev. Mod. Phys. 77, 633 (2005).
\bibitem{8} S. E. Harris, J. E. Field, A. Imamoglu, Nonlinear optical processes using electromagnetically induced transparency. Phys. Rev. Lett. 64, 1107 (1990).
\bibitem{9} B. K. Teo, D. Feldbaum, T. Cubel, J. R. Guest, P. R. Berman, G. Raithel, Autler-Townes spectroscopy of the 5S1/2-5P3/2-44D cascade of cold 85Rb atoms. Phys. Rev. A 68, 053407 (2003).
\bibitem{10} H. Zhang, L. M. Wang, J. Chen, S. X. Bao, L. J. Zhang, J. M. Zhao, S. T. Jia, Autler-Townes splitting of a cascade system in ultracold cesium Rydberg atoms. Phys. Rev. A 87, 033835 (2013).
\bibitem{11} T. Vogt, M. Viteau, J. M. Zhao, A. Chotia, D. Comparat, P. Pillet, Dipole blockade through Rydberg Forster resonance energy transfer. Phys. Rev. Lett. 97, 083003 (2006).
\bibitem{12}	W. H. Li, P. J. Tanner, T. F. Gallagher, Dipole-dipole excitation and ionization in an ultracold gas of Rydberg atoms. Phys. Rev. Lett. 94, 173001 (2005).
\bibitem{13} M. D. Lukin, M. Fleischhauer, R. C{\^o}t{\'e}, L. M. Duan, D. Jaksch, J. I. Cirac, P. Zoller, Dipole blockade and quantum information processing in mesoscopic atomic ensembles. Phys. Rev. Lett. 87, 037901 (2001).
\bibitem{14} A. Reinhard, C. T. Liebisch, K. C. Younge, P. R. Berman, G. Raithel, Rydberg-Rydberg collisions: resonant enhancement of state mixing and penning ionization. Phys. Rev. Lett. 100, 123007 (2008).
\bibitem{15} A. D. Bounds, N. C. Jackson, R. K. Hanley, R. Faoro, E. M. Bridge, P. Huillery, M. P. A. Jones, Rydberg-dressed magneto-optical trap. Phys. Rev. Lett. 120, 183401 (2018).
\bibitem{16} U. D. Rapol, V. Natarajan, Doppler-free spectroscopy in driven three-level systems. Eur. Phys. J. D 28, 317-322 (2004).
\bibitem{17} Y. H. Wang, H. J. Yang, Z. J. Du, T. C. Zhang, J. M. Wang, Autler-Townes doublet in novel sub-Doppler spectra with caesium vapour cell. Chinese Phys. 15, 138-142 (2006).
\bibitem{18} P. Thoumany, T. H{\"a}nsch, G. Stania, L. Urbonas, T. Becker, Optical spectroscopy of rubidium Rydberg atoms with a 297 nm frequency-doubled dye laser. Opt. Lett. 34, 1621-1623 (2009).
\bibitem{19} E. Urban, T. A. Johnson, T. Henage, L. Isenhower, D. D. Yavuz, T. G. Walker, M. Saffman, Observation of Rydberg blockade between two atoms. Nature Phys. 5, 110 (2009).
\bibitem{20} S. X. Bao, H. Zhang, J. Zhou, L. J. Zhang, J. M. Zhao, L. T. Xiao, S. T. Jia, Polarization spectra of Zeeman sublevels in Rydberg electromagnetically induced transparency. Phys. Rev. A 94, 043822 (2016).
\bibitem{21} H. Cheng, H. Wang, S. S. Zhang, P. P. Xin, J. Luo, H. P. Liu, High quality electromagnetically induced transparency spectroscopy of 87Rb in a buffer gas cell with a magnetic field. Chinese Phys. B 26, 074204 (2017).
\bibitem{22} A. M. Hankin, Rydberg excitation of single atoms for applications in quantum information and metrology. Ph. D Thesis, University of New Mexico (2014).
\bibitem{23} J. Y. Wang, J. D. Bai, J. He, J. M. Wang, Development and characterization of a 2.2 W narrow-linewidth 318.6 nm ultraviolet laser. J. Opt. Soc. Am. B 33, 2020-2025 (2016).
\bibitem{24} J. D. Bai, J. Y. Wang, J. He, J. M. Wang, Electronic sideband locking of a broadly tunable 318.6 nm ultraviolet laser to an ultrastable optical cavity. J. Opt. 19, 045501 (2017).
\bibitem{25} C. Wieman, T. W. Hanch, Doppler-free laser polarization spectroscopy. Phys. Rev. Lett. 36, 1170-1173 (1976).
\bibitem{26} D. A. Steck, Cesium D Line Data, available online at http://steck.us/alkalidata, 9-10 (2010).
\bibitem{27} G. S. Agarwal, Nature of the quantum interference in electromagnetic-field-induced control of absorption. Phys. Rev. A 55, 2467-2470 (1997).
\bibitem{28} S. Menon, G. S. Agarwal, Gain components in the Autler-Townes doublet from quantum interferences in decay channels. Phys. Rev. A 61, 013807 (1999)
\bibitem{29} P. Thoumany, T. H{\"a}nsch, G. Stania, L. Urbonas, T. Becker, Optical spectroscopy of rubidium Rydberg atoms with a 297 nm frequency-doubled dye laser. Opt. Lett. 34, 1621-1623 (2009).
\bibitem{30} T. Baluktsian, Rydberg interaction between thermal atoms: Van der Waals-type Rydberg-Rydberg interaction in a vapor cell experiment. Ph. D Thesis, University of Stuttgart (2013).
\bibitem{31} A. K. Mohapatra, T. R. Jackson, C. S. Adams, Coherent optical detection of highly excited Rydberg states using electromagnetically induced transparency. Phys. Rev. Lett. 98, 113003 (2007).
\bibitem{32} Y. C. Jiao, X. X. Han, Z. W. Yang, J. K. Li, G. Raithel, J. M. Zhao, S. T. Jia, Spectroscopy of cesium Rydberg atoms in strong radio-frequency fields. Phys. Rev. A 94, 023832 (2016).
\bibitem{33} A. Sargsyan, G. Hakhumyan, C. Leroy, Y. Pashayan-Leroy, A. Papoyan, D. Sarkisyan, M. Auzinsh, Hyperfine Paschen-Back regime in alkali metal atoms: consistency of two theoretical considerations and experiment. J. Opt. Soc. Am. B 31, 1046-1053 (2014).
\bibitem{34} S. X. Bao, W. G. Yang, H. Zhang, L. J. Zhang, J. M. Zhao, S. T. Jia, Splitting of an electromagnetically induced transparency window of a cascade system with 133Cs Rydberg atoms in a static magnetic field. J. Phys. Soc. Jpn. 84, 104301 (2015).
\bibitem{35} L. J. Zhang, S. X. Bao, H. Zhang, G. Raithel, J. M. Zhao, L. T. Xiao, S. T. Jia, Nonlinear Zeeman effect, line shapes and optical pumping in electromagnetically induced transparency. arXiv:1702.04842v1.
\bibitem{36} J. Y. Wang, J. D. Bai, J. He, J. M. Wang, Single-photon cesium Rydberg excitation spectroscopy using 318.6-nm UV laser and room-temperature vapor cell. Opt. Express 25, 22510-22518 (2017).


\end{thebibliography}
\end{document}